\documentclass[journal,transmag]{IEEEtran}
\usepackage{amsmath}
\usepackage{graphicx}
\usepackage[utf8]{inputenc}
\usepackage[english]{babel}

\begin{document}

\title{Colored-noise magnetization dynamics:\\ from weakly to strongly correlated noise}

\author{
\IEEEauthorblockN{
Julien Tranchida\IEEEauthorrefmark{1,2},
Pascal Thibaudeau\IEEEauthorrefmark{1},
Stam Nicolis\IEEEauthorrefmark{2}, 
}
\IEEEauthorblockA{\IEEEauthorrefmark{1}CEA/DAM/Le Ripault, BP 16, F-37260, Monts, France}
\IEEEauthorblockA{\IEEEauthorrefmark{2}CNRS-Laboratoire de Mathématiques et Physique Théorique (UMR 7350), \\
Fédération de Recherche "Denis Poisson" (FR2964), Département de Physique, Université de Tours, \\
Parc de Grandmont, F-37200, Tours, France}
\thanks{Corresponding author: J. Tranchida (email: julien.tranchida@cea.fr).}
}

%\markboth{IEE Trans. on Magnetics,~Vol.~13, No.~9, September~2014}
%{Shell \MakeLowercase{\textit{et al.}}: Bare Demo of IEEEtran.cls for Journals}

\IEEEtitleabstractindextext{
\begin{abstract}
Statistical averaging theorems allow us to derive a set of equations for the averaged magnetization dynamics in the presence of  colored 
(non-Markovian) noise. The non-Markovian character of the noise is described by a finite auto-correlation time, $\tau$, that can be identified with  the finite response time of the thermal bath to  the system of interest.
Hitherto, this model was only tested  for the case of weakly correlated noise (when $\tau$ is equivalent or smaller than the integration timestep).
In order to probe its validity for a broader range of auto-correlation times, a non-Markovian integration model, based on the stochastic 
Landau-Lifshitz-Gilbert equation is presented.
Comparisons between the two models are discussed, and these provide evidence  that both formalisms remain equivalent, even for strongly correlated noise (i.e. $\tau$ much larger than the integration timestep).
\end{abstract}
}

\maketitle
 %%%%%%%%%%%%%%%%%%%%%%%%%%%%%%%%%%%%%%%
\section{Introduction}
%%%%%%%%%%%%%%%%%%%%%%%%%%%%%%%%%%%%%%%
The study of the effects of thermal fluctuations  on the magnetization dynamics, by way of stochastic simulations, is a topic that has received much attention \cite{Brown-Jr:1963tp,coffey2012thermal}.
What has also been the subject  of numerous studies, in the field  of coarse-grained simulations \cite{thibaudeau2015frequency}, 
is the development of multiscale models, able to simulate the average magnetic behaviour of a spin assembly under thermal stresses \cite{garanin1990dynamics,kazantseva2008towards}.
But in both contexts, the coupling  between the thermal bath and the spin assembly has mainly been taken into account within the framework of the white-noise limit,
and few studies \cite{bose2010correlation,atxitia2009ultrafast} have  considered the effects of a finite auto--correlation time for the response of the thermal bath, i.e. the case of colored noise. 

However, the simulation of ultra-fast magnetic experiments~\cite{kirilyuk2010ultrafast,vahaplar2012all} is now able to probe time-scales of the dynamics 
of comparable order of magnitude~\cite{rossi2005dynamics,vaterlaus1991spin} to the  autocorrelation time of the external random field.
In these situations, a colored form for the noise, introducing a finite auto-correlation time, can provide a better  description for 
the random variables of the bath \cite{bose2010correlation,atxitia2009ultrafast}.

The bath variables can be defined by random vectors $\vec{\tilde{\omega}}$ whose components $\tilde{\omega}_i$ are drawn from a Gaussian distribution with zero mean and second-order autocorrelation functions: 
\begin{equation}
\langle \tilde{\omega}_i(t) \tilde{\omega}_j(t') \rangle=\frac{D}{\tau}\delta_{ij} \exp \left(-\frac{\lvert t-t' \rvert}{\tau} \right)
\label{Colorednoise}
\end{equation}
with $\tau$ the auto-correlation time, $D$  the amplitude of the noise.  Wick's theorem allows us to express  all  other autocorrelation functions in terms of these one- and  two-point functions~\cite{justin1989quantum}.

When the auto-correlation time vanishes ($\tau \rightarrow 0$), the well-known white-noise limit is recovered: 
$\langle \tilde{\omega}_i(t) \tilde{\omega}_j(t') \rangle=2D\delta_{ij} \delta \left( t-t' \right)$. 

In  previous work~\cite{tranchida2015closing}, statistical averaging theorems~\cite{shapiro1978formulae}
were applied to a non-Markovian form of the stochastic Landau-Lifshitz-Gilbert equation (sLLG), and they allowed us to derive a new hierarchy of 
equations for the simulation of averaged magnetization dynamics.
Under Gaussian assumptions, the  hierarchy of equations was closed, and a new model, presenting both transverse and longitudinal forms of 
damping, was obtained. This model is denoted as the dynamical Landau-Lifschitz-Bloch model (dLLB).
A first benchmark of tests was performed, and the validity of our closure assumptions was checked to hold  for weakly correlated noise (i.e. $\tau \rightarrow 0$), and with a constant applied magnetic field. However, the case of strongly correlated noise was left for future investigations--in the present paper we discuss our first results  along this path.

In section \ref{Section2}, the non-Markovian sLLG model is presented, and compared with other works based on non-Markovian stochastic magnetization dynamics. Then, some numerical experiments are presented in order to assess some aspects of this model, and to probe the impact of the auto-correlation time $\tau$ on the magnetization dynamics.

Section \ref{Section3} introduces our dLLB model. The results of numerical experiments, performed  
to assess the range of validity of our closure assumptions, for different domains of  values of $\tau$, i.e. beyond Markovian situations are discussed.

We finally conclude, in section~\ref{conclusion}, on the relevance  of our dLLB model, when the auto-correlation time for the bath variables is no longer negligible, and present some ideas about the situations where  our closure assumptions may break down  and, therefore,  when the validity of our model attains its limits. 
%%%%%%%%%%%%%%%%%%%%%%%%%%%%%%%%%%%%%%%
\section{Non-Markovian sLLG modelling}\label{Section2}
%%%%%%%%%%%%%%%%%%%%%%%%%%%%%%%%%%%%%%%
The Einstein summation convention is used throughout, with  Latin indices standing for vector components, $\epsilon_{ijk}$ is the anti-symmetric Levi-Civita pseudo-tensor 
and $s_i$ denotes the component of the normalized magnetization vector ($\left|\vec{s} \right|=1$). 
Up to a renormalization over the noise \cite{mayergoyz2009nonlinear}, the stochastic Landau-Lifshitz-Gilbert (sLLG) equation can be written as follows:
\begin{equation}
\frac{\partial s_i}{\partial t}=\frac{1}{1+\lambda^2} \epsilon_{ijk} s_k \left[\omega_j + \tilde{\omega}_j -\lambda\epsilon_{jlm} \omega_l s_m \right]
\label{sLLG}
\end{equation}
with $\omega_j$ the precession frequency assumed independent of the spin state and $\tilde{\omega}_j$ describes the (fluctuating) contribution  of the thermal bath.

When the $\vec{\tilde{\omega}}$ become non-Markovian random variables, drawn from  a Gaussian distribution, defined by the autocorrelation functions (\ref{Colorednoise}), the sLLG equation (eq.~\ref{sLLG}) can be recast into the following set of equations \cite{fox1988fast,berdichevsky1999stochastic}:
\begin{eqnarray}
\frac{\partial s_i}{\partial t}                 &=& \frac{1}{1+\lambda^2} \epsilon_{ijk} s_k \left[\omega_j + \tilde{\omega}_j -\lambda\epsilon_{jlm} \omega_l s_m \right] \label{sLLG_colored_eq} \\
\frac{\partial \tilde{\omega}_j}{\partial t}    &=& -\frac{1}{\tau} \left(\tilde{\omega}_j-\xi_j \right) \label{sLLG_colored_noise}
\end{eqnarray}
and where $\vec{\xi}$ are  new Gaussian random variables, such that: $\langle \xi_j\left( t\right) \rangle=0$ and $\langle \xi_i(t) \xi_j(t') \rangle=2D \delta_{ij} \delta\left( t-t'\right)$.

The bath variables are now, themselves, solutions  of a stochastic differential equation. This, indeed, is what introduces inertial effects, that are described by the auto-correlation time $\tau$. 

An interesting feature of this new set of equations is that, for the case of  no transverse damping ($\lambda=0$ in eq.~\ref{sLLG_colored_eq}), we immediately recover the formalism developed by Myasaki and Seki (MS) \cite{miyazaki1998brownian}. In this way, the set of equations (\ref{sLLG_colored_eq},\ref{sLLG_colored_noise}) can be seen as a generalization of their work in the case of finite transverse damping.

However, their equation on the bath variables takes a slightly different form than in our work, because they introduced a phenomenological form of damping, that depends on the magnetization state (and with a spin coupling constant $\chi$). 
The dynamics of $\vec{\tilde{\omega}}$ is indeed damped by a factor that depends on the value of the magnetization, and therefore, 
the corresponding distribution (for the bath variable) has no reason to remain Gaussian (especially in the case of a non-trivial Hamiltonian for the spin system).
In our case, by construction, it is ensured that the bath variables $\tilde{\omega}_i$ are, both, always drawn  from a Gaussian distribution, and remain sensitive to the inertial effects of the  memory kernel, defined by the color of the noise. 

This set of equations (\ref{sLLG_colored_eq}) and (\ref{sLLG_colored_noise}) is then integrated using a third-order Omelyan's
algorithm \cite{omelyan2003symplectic,ma2008large} that respects the symplectic properties of equation (\ref{sLLG_colored_eq}). Regarding the noise, this system is formally integrated as $\tilde{\omega}_i(t)\equiv e^{tL_{\tilde{\omega}}}\tilde{\omega}_i(0)=\int_0^te^{-(t-u)/\tau}\xi_i(u)du$ with $L_{\tilde{\omega}}$ is the Liouville operator corresponding to Eq.(\ref{sLLG_colored_noise}). Because the Liouville operators of the spin and the noise commute, an integration of the colored system is performed by the combination of the two operators as $e^{tL}(s_i(0),\tilde{\omega}_i(0))=e^{tL_s}e^{tL_{\tilde{\omega}}}(s_i(0),\tilde{\omega}_i(0))=(s_i(t),\tilde{\omega}_i(t))$.

It should be stressed that this scheme indeed incorporates  the Stratonovich {\em Ansatz}, whose theoretical foundations have been studied by Aron {\em et al.}~\cite{aron2014magnetization} for the case of the sLLG.

\begin{figure}[thp]
\centering
\resizebox{0.99\columnwidth}{!}{\includegraphics{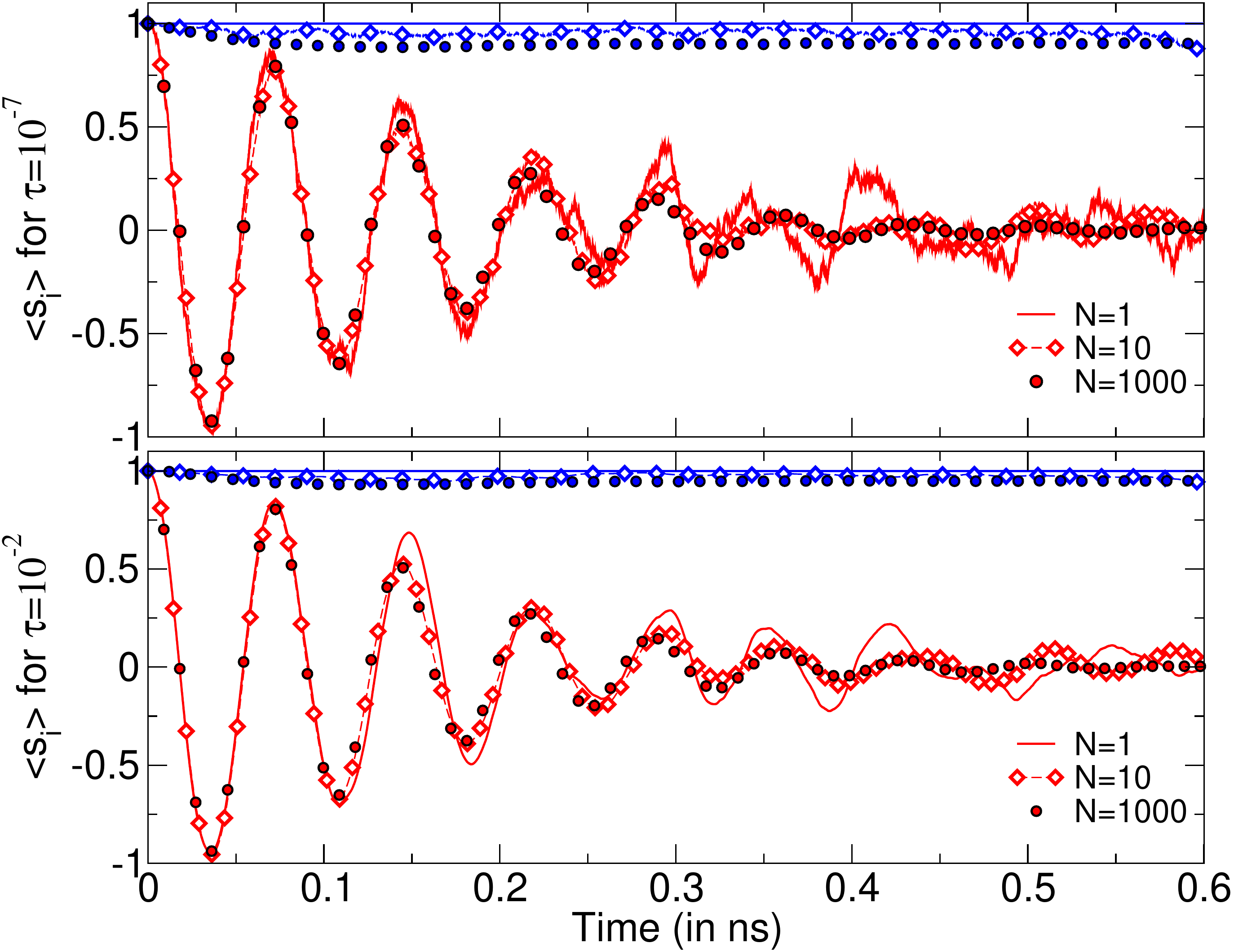}}
\caption{
Comparisons between simulations performed with the effective stochastic model, for two different values of the auto-correlation time:
$\tau=10^{-7}$ above, and  $\tau=10^{-2}$ below.
The red curves plot the dynamics of a component of the averaged magnetization vector ($\langle s_x\rangle$) in a constant magnetic field in the $z$-direction, and the blue curves plot the square of the averaged magnetization norm, for different repetition values N (N=1, 10 and 1000). Parameters of the simulations: \{$D=1$~rad.GHz; $\lambda=0.1$; $\vec{\omega}=(0,0,\omega_z)$, with $\omega_z=176$~rad.GHz; timestep $\Delta t=10^{-7}$ ns\}. Initial conditions: $\vec{s}(0)=\left(1,0,0 \right)$.   \label{Fig0}}
\end{figure}

An effective averaging model is then constructed: the integration of the set of stochastic equations (\ref{sLLG_colored_eq},\ref{sLLG_colored_noise}) 
is performed one thousand times, each time for a different realization of the random variables $\xi_i$. 
The results for the spin realizations are stored, and statistical averages thereof are computed. 
This methodology presents, of course, some disadvantages: first of all, that it is purely computational; also, that it requires significant storage for the configurations.  
However, these are more than compensated for by the fact that it provides an exact formulation and resolution of the Langevin-like stochastic equation (our non-Markovian sLLG equation). 
Fig.~\ref{Fig0} presents some results directly obtained with the effective averaging model.

Fig.~\ref{Fig0} highlights the following interesting features of our system:
A first, very important, result is that, for the case $N=1$, that plots two stochastic simulations, without any averaging effects,
increasing the auto-correlation time, $\tau$, leads to stronger inertial effects, that can be directly seen from the curves of $\langle s_x\rangle$. 
Indeed, from  fig.~\ref{Fig0}, we can deduce that this variation does not have a strong impact on the amplitude of the fluctuations of $\langle s_x\rangle$, but tends to smoothen them out.

Some other features can be observed, concerning the square of the averaged magnetization norm. 
First, when $N=1$, the norm of the magnetization remains perfectly constant and equal to unity (without any  renormalization necessary over the time steps). 
This is a good test of the validity of the integration scheme that has been applied, and therefore make all our procedures 
(eqs.~(\ref{sLLG_colored_eq}) and (\ref{sLLG_colored_noise}), and the integration scheme)  a good framework for non-Markovian atomistic simulations. 
Moreover, when the number $N$, of realizations increases (which is equivalent to a better approximation of the statistically averaged quantities), 
the square of the averaged magnetization norm stabilizes below unity.
This constitutes a simple and very direct recovery of the equivalence between the presence of a longitudinal form of damping, and statistical averaging in a thermalized magnetic system.

Therefore,  the behavior in time of the average magnetization dynamics (that our dLLB model is expected to track) can be used to realize  direct comparisons between the two models,  for the same set of parameters.
%%%%%%%%%%%%%%%%%%%%%%%%%%%%%%%%%%%%%%%
\section{Testing the limits of the auto-correlation time effects }\label{Section3}
%%%%%%%%%%%%%%%%%%%%%%%%%%%%%%%%%%%%%%%
The set of statistically averaged equations, called dLLB model, is now introduced.
With $\langle\cdot\rangle$ denoting the statistical average over the noise, one has the following set of $21$ equations:
\begin{eqnarray}
\frac{\partial \langle s_i\rangle}{\partial t}   &=&\frac{1}{1+\lambda^2} \left[ \epsilon_{ijk}\omega_j \langle s_k\rangle+ \epsilon_{ijk}\langle \tilde{\omega}_j s_k\rangle \right.\nonumber\\
                                 ~&~&-\left.\lambda \epsilon_{ijk}\epsilon_{jlm}\omega_l  \langle s_k s_m \rangle \right] \label{Averagedset1}\\
\frac{\partial \langle \tilde{\omega}_i s_j\rangle}{\partial t}&=&-\frac{1}{\tau} \langle \tilde{\omega}_i s_j\rangle+\frac{1}{1+\lambda^2} \Big[\epsilon_{jkl}\omega_k  \langle \tilde{\omega}_i s_l\rangle  \Big.  \nonumber\\
                                 ~&~&+\frac{D}{\tau}\epsilon_{jil}\langle s_l\rangle-\lambda\epsilon_{jkl}\epsilon_{kmn}\omega_m \left( \langle s_l \rangle\langle\tilde{\omega}_i s_n\rangle \right.\nonumber\\
                                 ~&~&+\Big.\left. \langle s_n\rangle\langle\tilde{\omega}_i s_l\rangle \right)\Big] \label{Averagedset2}\\
\frac{\partial \langle s_i s_j\rangle}{\partial t}&=&\frac{1}{1+\lambda^2} \Big[ \epsilon_{jkl}\omega_k \langle s_i s_l\rangle  \Big.\nonumber\\
                                 ~&~&+ \epsilon_{jkl} \left( \langle s_i\rangle \langle \tilde{\omega}_k s_l\rangle +\langle s_l\rangle  \langle \tilde{\omega}_k s_i\rangle  \right) \nonumber\\
                                 ~&~&- \lambda\epsilon_{jkl}\epsilon_{kmn}\omega_m \left( \langle s_i\rangle  \langle s_l s_n\rangle \right.\nonumber\\
                                 ~&~&+ \langle s_l\rangle \langle s_i s_n\rangle  +\langle s_ n\rangle \langle s_i s_l\rangle-\Big.\left.2\langle s_i\rangle\langle s_l\rangle\langle s_n\rangle\right)\Big]  \nonumber\\
                                 ~&~&+ \left(i \leftrightarrow j\right) \label{Averagedset3}
\label{AveragedSystem}
\end{eqnarray}
Equations (\ref{Averagedset1}) and (\ref{Averagedset3}) are strongly related to those derived by Garanin \emph{et al.} \cite{garanin1990dynamics}, 
and obtained under slightly different assumptions. Equation (\ref{Averagedset2}) is deduced from the application of the Shapiro-Loginov 
formula \cite{shapiro1978formulae}, and defines the time evolution of the mixed moments of the thermal bath variables and  the spin system variables.

This model was only shown to be valid for the simulation of averaged magnetization dynamics in interaction with a constant magnetic field, 
and when the auto-correlation time can be considered to be ``small'' (i.e. close to the Markovian limit) \cite{tranchida2015closing}. 
Under those conditions, the Gaussian assumptions that have been applied to the moments (and that allowed us to close the hierarchy of moments 
and to integrate our dLLB model) hold. 

\begin{figure}[thp]
\centering
\resizebox{0.99\columnwidth}{!}{\includegraphics{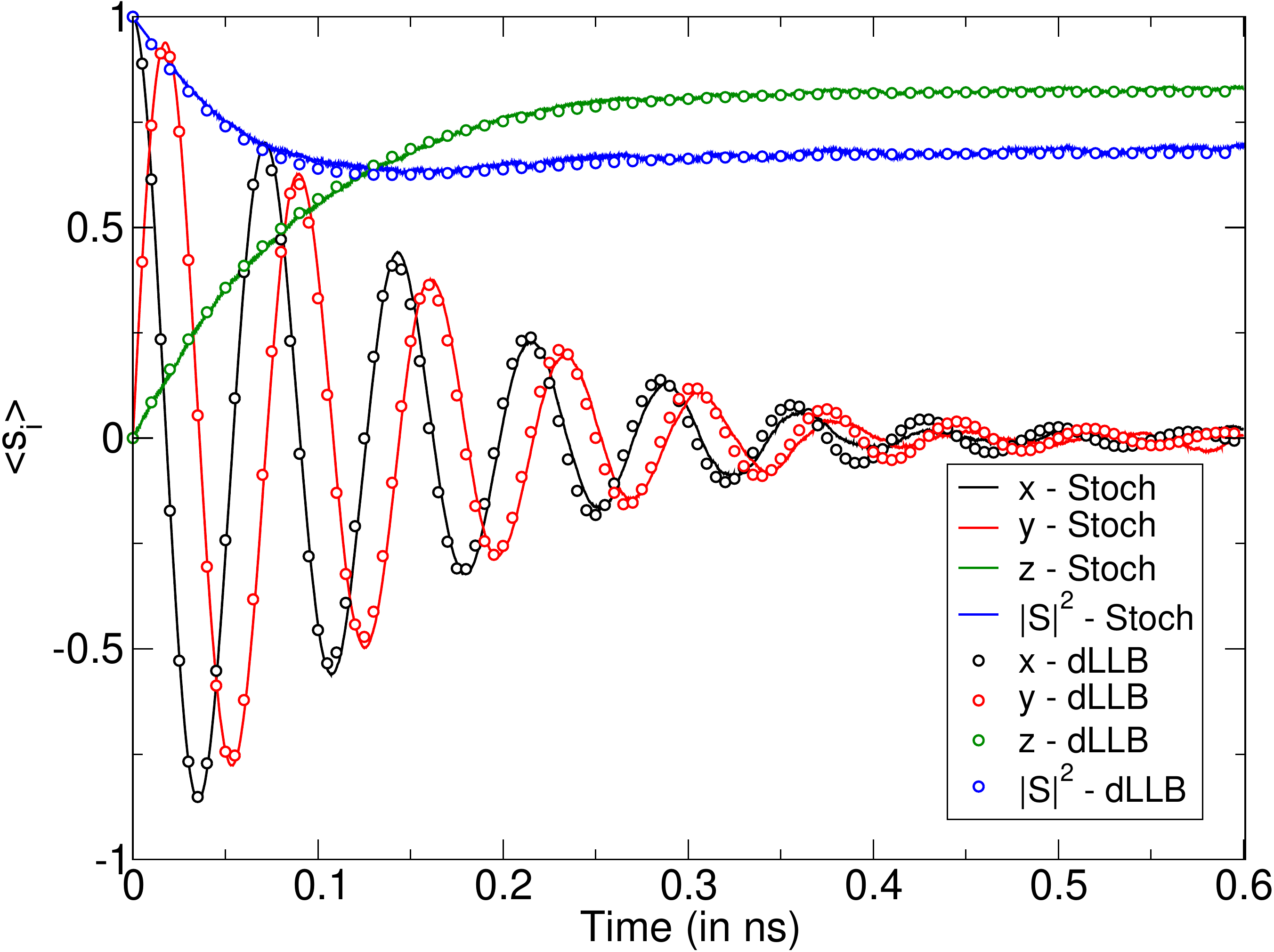}}
\caption{
Relaxation of the averaged magnetization dynamics in  a constant magnetic field in the $z$-direction for  ``weakly'' correlated noise, $\tau=10^{-7}$ ns. 
The solid lines plot the stochastic case ($10^3$ repetitions), dots are for the dLLB model. 
Parameters of the simulations: \{$D=1$~rad.GHz; $\lambda=0.1$; $\vec{\omega}=(0,0,\omega_z)$, with $\omega_z=176$~rad.GHz; timestep $\Delta t=10^{-7}$ ns\}. 
Initial conditions: $\vec{s}(0)=\left(1,0,0 \right)$.   \label{Fig1}}
\end{figure}

\begin{figure}[thp]
\centering
\resizebox{0.99\columnwidth}{!}{\includegraphics{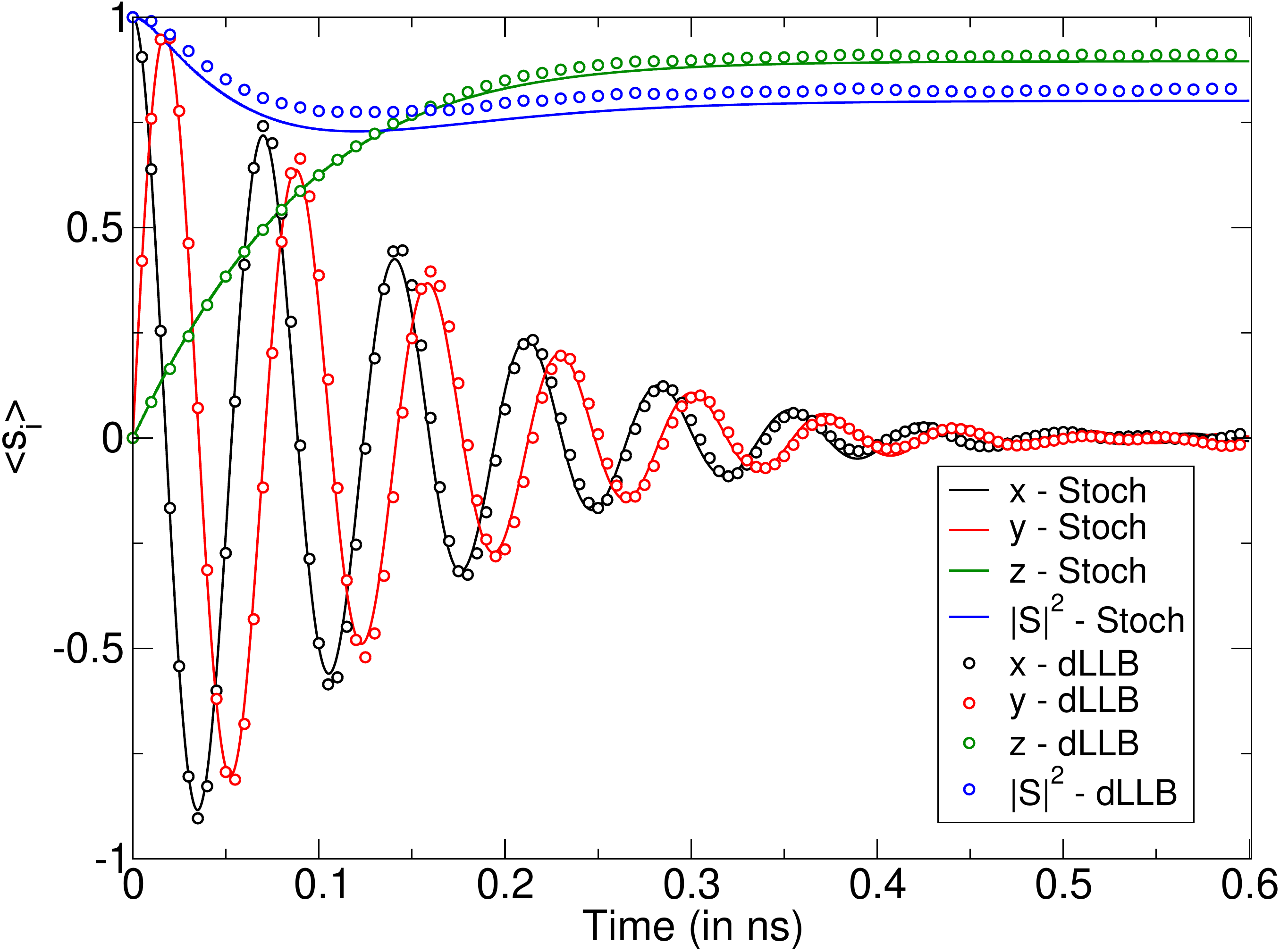}}
\caption{
Relaxation of the averaged magnetization dynamics in a constant magnetic field in the $z$-direction, for ``strongly'' correlated noise: $\tau=10^{-2}$ ns. All the other parameters and initialization setups are taken equal to those of the figure~\ref{Fig1}.\label{Fig2}}
\end{figure}

We now want to probe the range of validity of our closure assumptions for other values of the auto-correlation time $\tau$.
This  is possible via a direct comparison between our dLLB set of equations and the effective averaging model. Figures \ref{Fig1} and \ref{Fig2} plot these comparisons for two very different values of the auto-correlation time $\tau$. 

For weakly correlated noise ($\tau=\Delta t$), fig.~\ref{Fig1} shows an excellent agreement between the two models. The value of the auto-correlation time, $\tau$ was then increased, while keeping the value of the amplitude of the noise, $D$, fixed. As shown in fig.~\ref{Fig2}, the results of the two models only begin to diverge one from the other for ``strongly'' auto-correlated noise ($\tau = 10^{-2}$ns).

Several orders of magnitude separate the two values of the auto-correlation time $\tau$; nonetheless, we remark that our results provide evidence that our dLLB model and the related closure assumptions remain valid, also,  for strongly non-Markovian noise.

Both stochastic and deterministic curves of fig.~\ref{Fig1} and \ref{Fig2} are strongly smoothed by the averaging. Therefore, the difference in terms of inertial effects, 
between the strongly and slightly Markovian cases (when $\tau=10^{-2}$, or $\tau=10^{-7}$, respectively) is barely observable (unlike it was on fig.\ref{Fig0}, for the situations of small realization numbers, N=1 and 10). But an interesting feature, that is recovered from the comparison of the two graphs, is that  increasing  the auto-correlation time leads, on  average, to an enhancement of longitudinal damping effects. An equivalent result was already observed by Atxitia \emph{ et al.} \cite{atxitia2009ultrafast} in  atomistic non-Markovian simulations, performed with the Myasaki and Seki (MS) model.
%%%%%%%%%%%%%%%%%%%%%%%%%%%%%%%%%%%%%%
\section{Conclusion}\label{conclusion}
%%%%%%%%%%%%%%%%%%%%%%%%%%%%%%%%%%%%%%
An integration method for the non-Markovian sLLG equation was presented. This scheme allows us to construct  an effective averaging model that solves the stochastic Langevin-like equation, for a fixed realization of the non--Markovian noise process and construct the magnetization correlation functions from the many realizations that were drawn--the only limitation is storage for the configurations. 

A first set of numerical experiments was performed with this effective averaging model (results plotted on fig.~\ref{Fig0}). It allowed us to test the consistency of our integration scheme, and to probe the effects of the auto-correlation time $\tau$ on the magnetization dynamics. Inertial effects were clearly observed with the variations of $\tau$, and the influence of statistical averaging on the magnetization norm was also recovered (equivalence between statistical averaging and longitudinal form of damping in a magnetic system connected to a thermal bath).

Then, comparisons between the numerical results obtained from  this model and our previously derived dLLB set of equations were carried out. These simulations lead us to two main conclusions. First, it was shown that, even across large intervals of values  of the auto-correlation time $\tau$ (from weakly correlated noise, $\tau=10^{-7}$~ns, to clearly non-Markovian cases, $10^{-2}$~ns), the closure assumptions of our hierarchy of equations hold, and our dLLB model remains valid. The application of these techniques also allowed us to recover the result that increasing the auto-correlation time for the bath variables not only enhances their inertial effects, but can also lead to stronger longitudinal damping effects, that can be measured in  the average magnetization dynamics. 

In future work \cite{Thibaudeau2015}, it will be shown that what does affect the closure assumptions and the validity of our model are the magnetic interactions that are taken into account, for example, the contribution of a uniaxial anisotropy, or an exchange interaction.  

In this case, it was observed that cumulants of the spin and bath variables, assumed to be equal to zero under Gaussian assumptions, are no longer taking negligible values (like the third order cumulants of the spin, for example). Therefore, non--Gaussian closure assumptions are required, and a particular attention has to be taken for the treatment of the mixed (noise and spin) moments. The stochastic approach to chaotic dynamical systems~\cite{nicolis1998closing} may provide valuable insights to this end.

\section*{Acknowledgments}
JT acknowledges financial support through a joint doctoral fellowship ``Région Centre-CEA'' under the grant agreement number 00086667. 
%\bibliographystyle{IEEEtran}
%\bibliography{IEEEabrv,INTERMAG}

\begin{thebibliography}{10}
\providecommand{\url}[1]{#1}
\csname url@samestyle\endcsname
\providecommand{\newblock}{\relax}
\providecommand{\bibinfo}[2]{#2}
\providecommand{\BIBentrySTDinterwordspacing}{\spaceskip=0pt\relax}
\providecommand{\BIBentryALTinterwordstretchfactor}{4}
\providecommand{\BIBentryALTinterwordspacing}{\spaceskip=\fontdimen2\font plus
\BIBentryALTinterwordstretchfactor\fontdimen3\font minus
  \fontdimen4\font\relax}
\providecommand{\BIBforeignlanguage}[2]{{%
\expandafter\ifx\csname l@#1\endcsname\relax
\typeout{** WARNING: IEEEtran.bst: No hyphenation pattern has been}%
\typeout{** loaded for the language `#1'. Using the pattern for}%
\typeout{** the default language instead.}%
\else
\language=\csname l@#1\endcsname
\fi
#2}}
\providecommand{\BIBdecl}{\relax}
\BIBdecl

\bibitem{Brown-Jr:1963tp}
W.~F. Brown~Jr, ``Thermal fluctuations of a single-domain particle,'' \emph{J.
  Appl. Phys.}, vol.~34, no.~4, pp. 1319--1320, 1963.

\bibitem{coffey2012thermal}
W.~T. Coffey and Y.~P. Kalmykov, ``Thermal fluctuations of magnetic
  nanoparticles: Fifty years after {Brown},'' \emph{J. Appl. Phys.}, vol. 112,
  no.~12, p. 121301, 2012.

\bibitem{thibaudeau2015frequency}
P.~Thibaudeau and J.~Tranchida, ``Frequency-dependent effective permeability
  tensor of unsaturated polycrystalline ferrites,'' \emph{J. Appl. Phys.}, vol.
  118, no.~5, p. 053901, 2015.

\bibitem{garanin1990dynamics}
D.~A. Garanin, V.~V. Ishchenko, and L.~V. Panina, ``Dynamics of an ensemble of
  single-domain magnetic particles,'' \emph{Theor. Math. Phys.}, vol.~82,
  no.~2, pp. 169--179, 1990.

\bibitem{kazantseva2008towards}
N.~Kazantseva, D.~Hinzke, U.~Nowak, R.~W. Chantrell, U.~Atxitia, and
  O.~Chubykalo-Fesenko, ``Towards multiscale modeling of magnetic materials:
  Simulations of {F}e{P}t,'' \emph{Phys. Rev. B}, vol.~77, no.~18, p. 184428,
  2008.

\bibitem{bose2010correlation}
T.~Bose and S.~Trimper, ``Correlation effects in the stochastic
  {L}andau-{L}ifshitz-{G}ilbert equation,'' \emph{Phys. Rev. B}, vol.~81,
  no.~10, p. 104413, 2010.

\bibitem{atxitia2009ultrafast}
U.~Atxitia, O.~Chubykalo-Fesenko, R.~W. Chantrell, U.~Nowak, and A.~Rebei,
  ``Ultrafast spin dynamics: the effect of colored noise,'' \emph{Phys. Rev.
  Lett.}, vol. 102, no.~5, p. 057203, 2009.

\bibitem{kirilyuk2010ultrafast}
A.~Kirilyuk, A.~V. Kimel, and T.~Rasing, ``Ultrafast optical manipulation of
  magnetic order,'' \emph{Rev. Mod. Phys.}, vol.~82, no.~3, p. 2731, 2010.

\bibitem{vahaplar2012all}
K.~Vahaplar, A.~M. Kalashnikova, A.~V. Kimel, S.~Gerlach, D.~Hinzke, U.~Nowak,
  R.~Chantrell, A.~Tsukamoto, A.~Itoh, A.~Kirilyuk \emph{et~al.}, ``All-optical
  magnetization reversal by circularly polarized laser pulses: Experiment and
  multiscale modeling,'' \emph{Phys. Rev. B}, vol.~85, no.~10, p. 104402, 2012.

\bibitem{rossi2005dynamics}
E.~Rossi, O.~G. Heinonen, and A.~H. MacDonald, ``Dynamics of magnetization
  coupled to a thermal bath of elastic modes,'' \emph{Phys. Rev. B}, vol.~72,
  no.~17, p. 174412, 2005.

\bibitem{vaterlaus1991spin}
A.~Vaterlaus, T.~Beutler, and F.~Meier, ``Spin-lattice relaxation time of
  ferromagnetic gadolinium determined with time-resolved spin-polarized
  photoemission,'' \emph{Phys. Rev. Lett.}, vol.~67, no.~23, p. 3314, 1991.

\bibitem{justin1989quantum}
J.~Zinn-Justin, ``Quantum field theory and critical phenomena,''
  \emph{Clarendon, Oxford}, 1989.

\bibitem{tranchida2015closing}
J.~Tranchida, P.~Thibaudeau, and S.~Nicolis, ``Closing the hierarchy for
  non-markovian magnetization dynamics,'' \emph{Physica B: Condensed Matter},
  2015.

\bibitem{shapiro1978formulae}
V.~E. Shapiro and V.~M. Loginov, ``Formulae of differentiation and their use
  for solving stochastic equations,'' \emph{Physica A: Statistical Mechanics
  and its Applications}, vol.~91, no.~3, pp. 563--574, 1978.

\bibitem{mayergoyz2009nonlinear}
I.~D. Mayergoyz, G.~Bertotti, and C.~Serpico, \emph{Nonlinear magnetization
  dynamics in nanosystems}.\hskip 1em plus 0.5em minus 0.4em\relax Elsevier,
  2009.

\bibitem{fox1988fast}
R.~F. Fox, I.~R. Gatland, R.~Roy, and G.~Vemuri, ``Fast, accurate algorithm for
  numerical simulation of exponentially correlated colored noise,'' \emph{Phys.
  Rev. A}, vol.~38, no.~11, p. 5938, 1988.

\bibitem{berdichevsky1999stochastic}
V.~Berdichevsky and M.~Gitterman, ``Stochastic resonance in linear systems
  subject to multiplicative and additive noise,'' \emph{Phys. Rev. E}, vol.~60,
  no.~2, p. 1494, 1999.

\bibitem{miyazaki1998brownian}
K.~Miyazaki and K.~Seki, ``Brownian motion of spins revisited,'' \emph{The
  Journal of chemical physics}, vol. 108, no.~17, pp. 7052--7059, 1998.

\bibitem{omelyan2003symplectic}
I.~P. Omelyan, I.~M. Mryglod, and R.~Folk, ``Symplectic analytically integrable
  decomposition algorithms: classification, derivation, and application to
  molecular dynamics, quantum and celestial mechanics simulations,''
  \emph{Comp. Phys. Comm.}, vol. 151, no.~3, pp. 272--314, 2003.

\bibitem{ma2008large}
P.-W. Ma, C.~H. Woo, and S.~L. Dudarev, ``Large-scale simulation of the
  spin-lattice dynamics in ferromagnetic iron,'' \emph{Phys. Rev. B}, vol.~78,
  no.~2, p. 024434, 2008.

\bibitem{aron2014magnetization}
C.~Aron, D.~G. Barci, L.~F. Cugliandolo, Z.~G. Arenas, and G.~S. Lozano,
  ``Magnetization dynamics: path-integral formalism for the stochastic
  {L}andau--{L}ifshitz--{G}ilbert equation,'' \emph{J. of Stat. Mech.: Theory
  and Experiment}, vol. 2014, no.~9, p. P09008, 2014.

\bibitem{Thibaudeau2015}
P.~Thibaudeau, J.~Tranchida, and S.~Nicolis, ``Non-markovian magnetization
  dynamics for uniaxial nanomagnets,'' \emph{Work in progress}, 2015.

\bibitem{nicolis1998closing}
C.~Nicolis and G.~Nicolis, ``Closing the hierarchy of moment equations in
  nonlinear dynamical systems,'' \emph{Phys. Rev. E}, vol.~58, no.~4, p. 4391,
  1998.

\end{thebibliography}

\end{document}